\documentclass[aps,pra,showpacs,twocolumn]{revtex4-1}
\usepackage{graphicx,latexsym,amssymb}
\usepackage{amsmath}
\usepackage{amsfonts}
\usepackage{times}
\usepackage[colorlinks={true}]{hyperref}
\hypersetup{citecolor={blue}, filecolor={blue}, linkcolor={blue}, urlcolor={blue}}

\begin{document}

\title{Invariant entanglement and generation of quantum correlations under global dephasing}

\author{G\"{o}ktu\u{g} Karpat}
\email{goktug.karpat@ieu.edu.tr}
\affiliation{Faculty of Arts and Sciences, Department of Physics, Izmir University of Economics, \.{I}zmir, 35330, Turkey}


\begin{abstract}
We investigate the dynamics of quantum entanglement and more general quantum correlations quantified respectively via negativity and local quantum uncertainty for two qubit systems undergoing Markovian collective dephasing. Focusing on a two-parameter family of initial two-qubit density matrices, we study the relation of the emergence of the curious phenomenon of time-invariant entanglement and the dynamical behavior of local quantum uncertainty. Developing an illustrative geometric approach, we demonstrate the existence of distinct regions of quantum entanglement for the considered initial states and identify the region that allows for completely frozen entanglement throughout the dynamics, accompanied by generation of local quantum uncertainty. Furthermore, we present a systematic analysis of different dynamical behaviors of local quantum uncertainty such as its sudden change or smooth amplification, in relation with the dynamics of entanglement.
\end{abstract}

\maketitle

\section{Introduction}

Quantum physics allows for the existence of correlations among physical systems, which are purely of quantum mechanical origin. Such genuinely quantum correlations have no analogue in classical physics and thus can only be formed among the constituents of composite quantum systems. The concept of quantum entanglement, which has been considered as the characteristic trait of quantum theory by Schr\"{o}dinger himself, is a paramount example of quantum correlations \cite{entrev}. Although it has been originally regarded as a mere philosophical issue of the quantum theory for a long time until few decades ago, it is now well accepted that entanglement lies at the heart of quantum information science, being a very fundamental resource for many useful applications of the quantum theory, such as quantum information processing, quantum cryptography and quantum computation \cite{book}.

Despite the fact that quantum entanglement has been considered to be the main resource of quantum information science, it has been discovered that, in certain applications, even quantum systems in separable states can achieve superior performance as compared to the classical approach to the same task \cite{sep}. Consequently, through various investigations in the recent literature, it has been shown that there might indeed exist different types of genuine quantum correlations responsible for this efficiency, which are more general than entanglement. That is, even separable quantum states with no entanglement can exhibit such general quantum correlations as long as the considered quantum system is in a mixed state.  Even though the introduction of quantum discord has paved the way for quantifying general type of correlations, there are now numerous discord-like measures in the literature \cite{discrev}. One such recently proposed measure of quantum correlations, known as local quantum uncertainty, is constructed upon the concept of skew information \cite{WYSI} and is intimately related to the quantification of coherence in quantum states \cite{WYSIqc}.

In practice, quantum systems are unavoidably open to interaction with their surrounding environment. As a result of this interaction, the principal open quantum system of interest gets coupled with its environment and therefore tends to lose its characteristic quantum traits, such as coherence and quantum correlations in composite systems, due to the destructive effects of the decoherence process \cite{bp}. Since success rates of quantum information tasks crucially depend on the existence of such critical quantum traits, it is of great importance to retaliate or avoid the detrimental effects of decoherence in realistic conditions for the realization of quantum technologies.

Dynamical behaviour of quantum correlations under different types of decoherece models has been the subject of intense research in recent years. It has been demonstrated that quantum states can suffer from the complete loss of entanglement in finite time, i.e., sudden death of entanglement \cite{sd}. On the other hand, discord-like general quantum correlations are more robust against noise and do not suffer from the phenomenon of sudden death \cite{robust}. Another striking phenomenon, known as the transition from classical to quantum decoherence, has been shown to occur under suitable local noise models for various discord-like correlations, including local quantum uncertainty \cite{frezunc,sabri,discrev2,discrev}. In this case, quantum correlations remarkably remain completely unaffected by the noisy time evolution for a finite time interval before beginning to decay asymptotically. Unlike discord-like correlations, quantum entanglement has not been reported to display such finite time freezing behaviour. However, it has been very recently demonstrated both theoretically \cite{theo} and experimentally \cite{exp} that, under a global dephasing noise setting, entanglement can freeze forever throughout the dynamics despite the decoherent evolution of the open system, a curious phenomenon known as the time-invariant entanglement. Until this date, the time-invariant nature of entanglement has not been investigated in relation with the decoherent dynamics of more general quantum correlations than entanglement such as the local quantum uncertainty. With this study, our main aim is to comparatively explore this relation in a global dephasing scenario.

In this work, we analyze the time-evolution of quantum entanglement and local quantum uncertainty for quantum systems composed of two-qubits undergoing decoherence. In particular, we consider quantum systems which can be initially described by a two-parameter family of X-shaped density matrices, and study the dynamics of entanglement and local quantum uncertainty under the effect of a global pure dephasing noise. We present an illustrative geometric approach to identify the different regions of entanglement for the considered family of states. This in turn enables us to explore for the first time the relation of the emergence of time-invariant entanglement (or sudden death of entanglement) to various dynamical behaviors of local quantum uncertainty, such as its generation, sudden change or loss during the dynamics.

\section{Global Dephasing Noise}

Let us firstly introduce the decoherence model that we intend to consider in our analysis. We suppose that a quantum system consisting of two qubits interacts with a common dephasing reservoir, where the role of the environment is played by stochastic fluctuations on qubits. The model Hamiltonian to describe such a scenario can be written as \cite{yueb}
\begin{equation}
H(t)= -\frac{1}{2} \xi(t)(\sigma_{z}^A \otimes \mathbb{I}^B +\mathbb{I}^A \otimes \sigma_{z}^B),
\end{equation}
where $\sigma_{z}$ is the Pauli operator in z-direction, $\mathbb{I}$ is the $2 \times 2$ identity matrix. Here, $\xi(t)$ is a stochastic noise parameter that satisfies the Markov noise conditions
\begin{align}
\langle \xi(t) \rangle =& 0,  \nonumber \\
\langle \xi(t)\xi(t') \rangle =& \Gamma \delta(t-t'),
\end{align}
with $ \langle \cdots \rangle $ being the ensemble average, and $\Gamma$ the damping rate associated with the noise parameter. Due to the Markovian nature of the noise, this model leads to a memoryless quantum evolution. Despite its simplicity, such a model can be regarded as the representative of a class of interactions generating a global pure dephasing process.

The time evolution of the density matrix describing the two-qubit system can be evaluated in a straightforward way as
\begin{equation}
\rho(t)= \langle U(t) \rho(0) U^{\dagger}(t) \rangle,
\end{equation}
where the ensemble averages are calculated for the noise parameter $\xi(t)$, and the time evolution operators reads
\begin{equation}
U(t)= \exp \left[-i \int_0^t \! dt' H(t') \ \right].
\end{equation}
Therefore, in the basis $\lbrace \vert ij\rangle : i=0, 1, j= 0, 1 \rbrace$, the dynamics of the two-qubit system can be expressed as
\begin{equation} \label{dyn}
\rho(t)=
\begin{pmatrix}
\rho_{11} & \rho_{12}\gamma & \rho_{13}\gamma & \rho_{14}\gamma^{4} \\
\rho_{21}\gamma & \rho_{22} & \rho_{23} & \rho_{24}\gamma \\
\rho_{31}\gamma & \rho_{32} & \rho_{33} & \rho_{34}\gamma \\
\rho_{41}\gamma^{4} & \rho_{42}\gamma & \rho_{43}\gamma & \rho_{44} \\
\end{pmatrix}
\end{equation}
where $\rho_{ij}$ represents the elements of the initial state density matrix, and decay rate is given by $\gamma(t)= e^{ -t \Gamma /2}$. We can note the existence of a decoherence free subspace which is a typical feature of global noise settings.
\vspace{-0.2cm}

\section{Quantum Correlations}

In this section, we introduce the correlation measures that we utilize to quantify entanglement and more general quantum correlations. We choose negativity as a measure to quantify entanglement due to its convenience for our purposes. For a given bipartite density matrix $\rho$, negativity is evaluated as two times the absolute sum of the negative eigenvalues of partial transpose of $\rho$ with respect to one of the subsystems \cite{neg}:
\begin{equation}
N(\rho)=\sum_{i}|\eta_{i}|-\eta_{i},
\end{equation}
where $\eta_{i}$ are all of the eigenvalues of $\rho^{T_{A}}$. That is, only the negative eigenvalues of the partially transposed density matrix actually contribute to the entanglement of the state. In other words, the considered state has vanishing entanglement if and only if all of the eigenvalues $\eta_i$ are non-negative.

In order to define the local quantum uncertainty to measure the more general genuine quantum correlations, we first need to introduce the concept of skew information, first proposed by Wigner and Yanase \cite{WYSI} as
\begin{equation} \label{WYSI}
I(\rho, K)=-\frac{1}{2}\textmd{Tr}[\sqrt{\rho},K]^2,
\end{equation}
where $[.,.]$ denotes the commutator of the matrices, $\rho$ is the density matrix describing the state of the considered quantum system, and $K$ is a quantum observable. We note that the skew information simplifies into the variance $V(\rho,K)=\textmd{Tr}\rho K^2-(\textmd{Tr} \rho K)^2$ for pure quantum states.

Wigner-Yanase skew information has been recently discussed in the context of the quantification of coherence possessed by a quantum state \cite{cohh}. It has been shown that although it does not satisfy all of the conditions for proper coherence monotones introduced in the framework presented in Ref. \cite{cohmon}, it can still be regarded as a quantifier of coherence in a different sense. That is, it can be regarded as a measure of asymmetry relative to the group of translations generated by the observable $K$, which is then interpreted as a quantifier of coherence for the state $\rho$ relative to the eigenbasis of the observable $K$ \cite{coh}. With this information at hand, it is straightforward to define local quantum coherence as $I(\rho_{AB}, K_{A}\otimes \mathbb{I}_{B})$, which measures the coherence contained in the first subsystem locally. We also note here that as the bipartite quantum systems we study in this work remain invariant upon swapping two qubits, their local quantum coherence is unchanged, independently of the measured subsystem.

Local quantum uncertainty as a measure of discord-like quantum correlations is based on the concept of local quantum coherence which we defined above. Indeed, local quantum uncertainty is nothing other than the optimized version of local quantum coherence over the set of local observables having the same (non-degenerate) spectrum, that is,
\begin{equation} 
U_A^\Gamma=\min_{K_A^\Gamma}I(\rho,K_A^\Gamma \otimes \mathbb{I}_{B} ),
\end{equation}
where $\Gamma$ denotes the spectrum of $K_A^\Gamma$, and the minimization over a chosen spectrum of observables gives a specific measure. To put it differently, there is in fact a family of discord-like measures that can be obtained from this definition for each $\Gamma$. Nevertheless, all members of this measure family can be shown to be equivalent in case of two-qubit states, in which case an analytical expression is fortunately available. Therefore, local quantum uncertainty reads \cite{WYSIqc}
\begin{equation} \label{lqu}
LQU=U_A(\rho_{AB})=1-\lambda_{\max}\{W_{AB}\},
\end{equation}
where $\lambda_{\max}$ is the maximum eigenvalue of the $3\times 3$ symmetric matrix $W_{AB}$ whose elements are computed as
\begin{equation}  \nonumber
(W_{AB})_{ij}=\textmd{Tr}\left\{\sqrt{\rho_{AB}}(\sigma_{iA}\otimes \mathbb{I}_{B})\sqrt{\rho_{AB}}(\sigma_{jA}\otimes \mathbb{I}_{B})\right\},
\end{equation}
where indices $i,j=\{x,y,z\}$ denote the Pauli operators. 

\section{Main Results}

Having briefly discussed the decoherence model and quantum correlations that are to be investigated in our study, we can now introduce the two-parameter density matrix family describing the two-qubit system. Throughout this work, we restrict our attention to the following X-shaped states:
\begin{equation} \label{inista}
\rho=\frac{1}{2}
\begin{pmatrix}
2r & 0 & 0& 2s \\
0 & 1-2r & 1-2r& 0 \\
0 & 1-2r & 1-2r& 0 \\ 
2s & 0 & 0& 2r \\
\end{pmatrix},
\end{equation}
where $r$ and $s$ are real valued parameters and the eigenvalues of the above density matrix are given by $\lambda_1=0$, $\lambda_2=1-2r$, $\lambda_3=r-s$, $\lambda_4=r+s$. Due to the condition that the eigenvalues of density matrices describing valid quantum states must be non-negative, the state parameters $r$ and $s$ can only assume values from the intervals $r\in[0,0.5]$ and $s\in[-0.5,0.5]$. The reason that we choose to analyse this family of states is two fold. First, such X-shaped states typically appear naturally in physical processes. Second, our specific choice of the two-parameter X-shaped state family in Eq. (\ref{inista}) allows for an analytical and illustrative geometric analysis of the problem.

Let us commence our investigation of the correlations with quantum entanglement quantified through negativity. Evaluating the partial transposition of the initial state density matrix in (\ref{inista}) with respect to any of the subsystems and computing the eigenvalues of the partially transposed matrix, we obtain
\begin{align}
\eta_1&= 1/2, &  \eta_2&=2 r-1/2, \nonumber \\ 
\eta_3&= 1/2 - r - s, &  \eta_4&=1/2 - r + s.
\end{align}
Since the first eigenvalue is positive and independent of time, only the last three eigenvalues can assume negative values depending on the state parameters. A simple observation is that only one of these last three eigenvalues can take a negative value for a given pair of state parameters $r$ and $s$. In other words, entanglement of the family given in Eq. (\ref{inista}) is only determined by one of the three eigenvalues $\eta_2$, $\eta_3$ and $\eta_4$. 

\begin{figure}[t]
\includegraphics[width=0.49\textwidth]{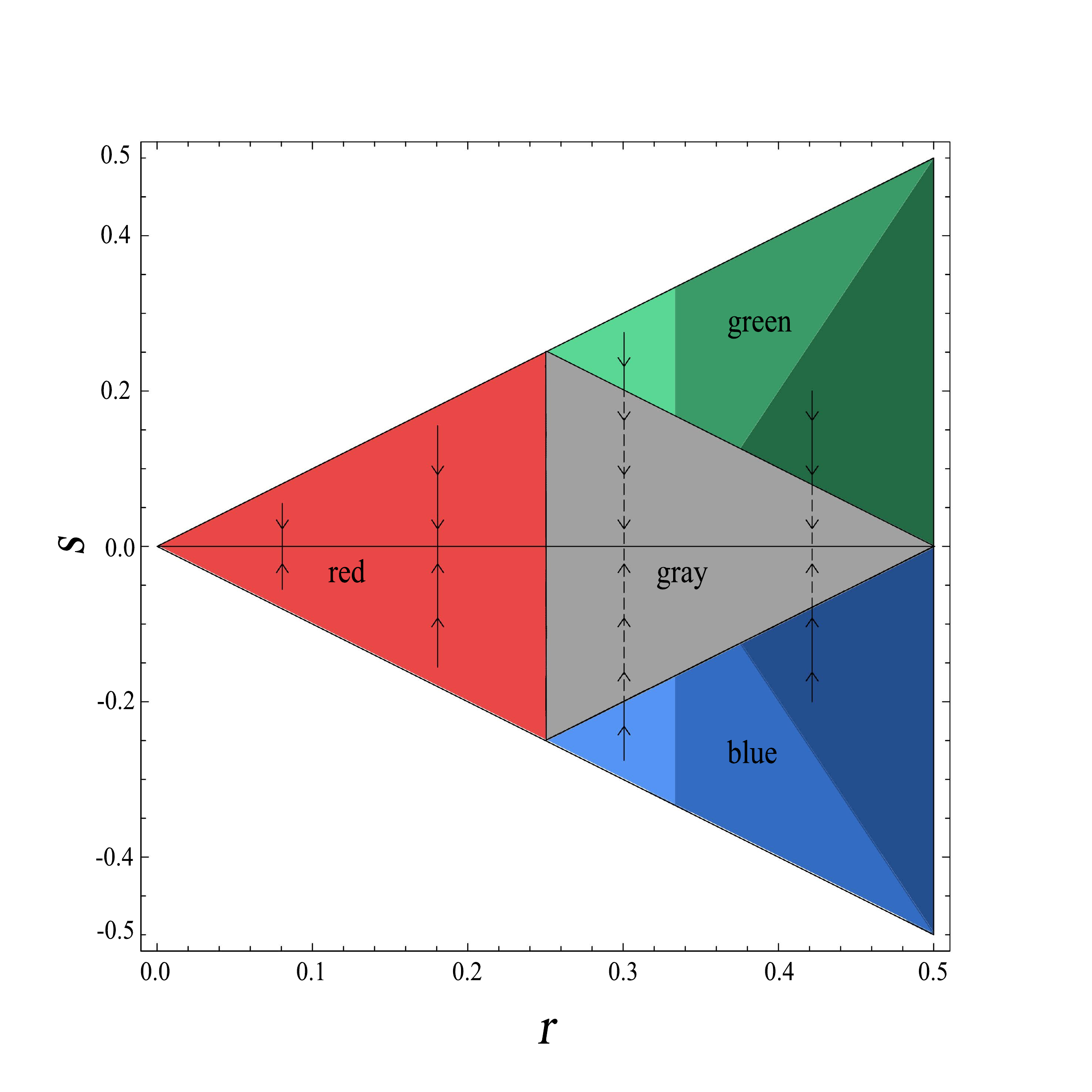}
\caption{Geometry of quantum entanglement for the considered two-parameter family of density matrices. While the red, green and blue triangular regions of entanglement are defined by the negativeness of the eigenvalues $\eta_2$, $\eta_3$ and $\eta_4$, respectively, the gray triangular region in the middle contains the separable states. Vertical black lines show few of the possible paths for the time evolution of the initial states. As solid black lines represent the path of the dynamics in entangled regions, dashed black lines denote the evolution in separable regions.}
\label{fig1}
\end{figure}

Since the class of states we study only depend on two real parameters, we can visualize the different regions of entanglement. In Fig. \ref{fig1}, we show the geometry of entangled regions. While the white regions are forbidden by the non-negativity of the eigenvalues $\lambda_i$, we observe that negativeness of the three eigenvalues of the partially transposed density matrix $\eta_i$ separate the coloured big triangle into four triangular regions. In particular, while separable states reside in the inner gray triangular region, the regions where $\eta_2$, $\eta_3$ and $\eta_4$ assume negative values correspond to the three remaining triangular regions coloured in red, green and blue, respectively. We also note that three of the Bell states are situated at the extreme points of these three triangles, namely at positions $r=0, s=0$ and $r=0.5, s=0.5$ and $r=0.5, s=-0.5$. Therefore, in each of the three regions, closer an arbitrary state gets to the extreme point, the more entangled it becomes. 

We now turn our attention to the dynamics of the considered density matrix family under global dephasing noise. Comparing Eq. (\ref{dyn}) to Eq. (\ref{inista}), we can simply observe that the time evolution does not change the general form of the initial density matrix and transforms its parameters as
\begin{equation} \label{time}
r(t)=r, \qquad s(t)=s\gamma^4(t).
\end{equation}
Thus, while the parameter $r$ remains unchanged during the decoherent dynamics thanks to the existence of the decoherence free subspace, the parameter $s$ decays and asymptotically vanishes as $t\rightarrow\infty$, i.e., the state loses some part of its coherence. This clearly demonstrates that any given initial state in Eq. (\ref{inista}) will evolve towards the $s=0$ line in Fig. \ref{fig1} without changing its $r$ parameter, which in turn implies that the evolution path of considered quantum states under global dephasing can be basically represented by vertical lines as shown in the figure. Focusing on the first region of entanglement shown by the red triangular region, we recall that the negative eigenvalue is given by $\eta_2=2 r-1/2$. Therefore, as this eigenvalue does not depend on $s$, it is easy to see that if the initial state is in this region, then its entanglement will remain forever frozen during the dynamics. That is, such initial states will follow the path of a vertical line towards the $s=0$ line on which entanglement is constant. Next, we investigate what happens if the initial state starts in one of the remaining entanglement regions depicted by green and blue triangles. In fact, let us just look at the green region defined by the negativeness of the eigenvalue $\eta_3= 1/2 - r - s$, since the blue region defined by $\eta_3= 1/2 - r + s$ would give similar results due to the symmetry of the problem about $s=0$ line. Provided that the initial state is chosen from the green triangular region, we see that it will cross into the gray separable region in finite time rather than asymptotically, that is, the state will suffer sudden death of entanglement losing all of its negativity. The time of sudden death can be calculated as 
\begin{equation}
t_{sd}=-\frac{1}{2\Gamma} \ln\left[\frac{1-2r}{2s}\right].
\end{equation}
The only exception occurs when the initial state is on $r=0.5$ line, since in this case the state cannot reach the gray separable region in finite time but rather asymptotically when $t\rightarrow\infty$. To sum up, as the phenomenon of time-invariant entanglement manifests for the initial states in the red coloured entanglement region, the initial states from the green and blue regions suffer from the sudden death of entanglement.

\begin{figure}[t]
\includegraphics[width=0.5\textwidth]{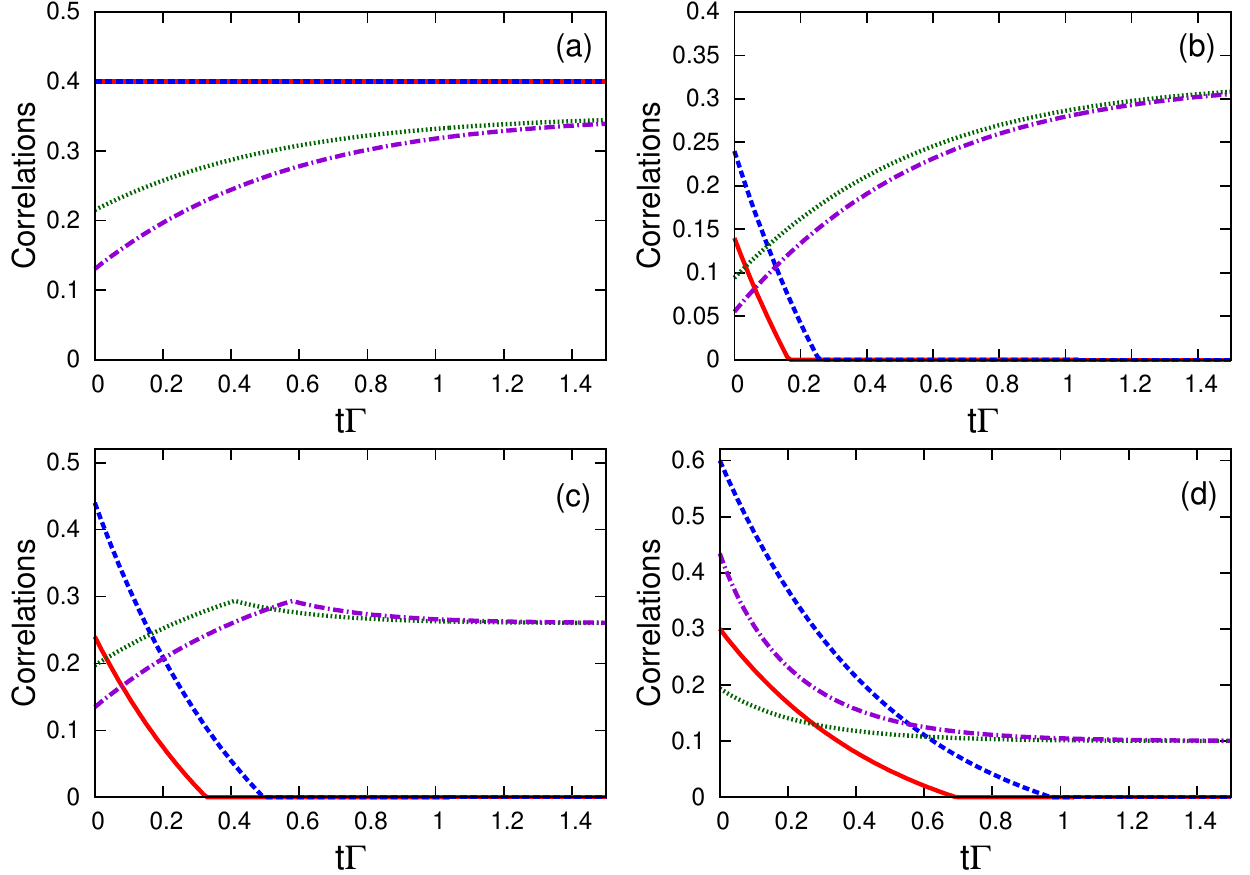}
\caption{Time evolution of the entanglement and more general discord-like correlations quantified respectively by negativity and and local quantum uncertainty for a pair of initial states. While entanglement is denoted by red solid and blue dashed lines, local quantum uncertainty is displayed by green dotted and purple dash-dotted lines in all insets (a-d). In (a) we have $r=0.15,s=0.07$ (red solid, green dotted) and $r=0.15,s=0.12$ (blue dashed, purple dash-dotted) for the initial states, in (b) we have $r=0.32,s=0.25$ (red solid, green dotted)  and $r=0.32,s=0.3$ (blue dashed, purple dash-dotted), in (c) we have $r=0.37,s=0.25$ (red solid, green dotted) and $r=0.37,s=0.35$ (blue dashed, purple dash-dotted), in (d) we have $r=0.45,s=0.2$ (red solid, green dotted) and $r=0.45,s=0.35$ (blue dashed, purple dash-dotted).}
\label{fig2}
\end{figure}

Next, we study the dynamics of general quantum correlations as quantified by local quantum uncertainty in relation with the entanglement regions depicted in Fig. \ref{fig1}. For initial state family given in Eq. (\ref{inista}), the analytical expression for the local quantum uncertainty can be calculated with the help of Eq. (\ref{lqu}) in a straightforward way as
\begin{equation}
LQU(\rho)=1-2\max[\beta_1,\beta_2,\beta_3],
\end{equation} 
where the individual terms $\beta_i$ above are given by
\begin{gather}
\beta_1=\sqrt{1-2r}\sqrt{r-s}, \quad \beta_2=\sqrt{1-2r}\sqrt{r+s}, \nonumber  \\
\beta_3=\sqrt{r-s}\sqrt{r+s}.
\end{gather}
We can now evaluate the dynamics of local quantum uncertainty recalling the evolution of the state parameters given in Eq. (\ref{time}) under the considered collective dephasing model.

At this point, let us once again focus on the upper part of Fig. \ref{fig1} where $s>0$ since our results will apply to the lower part automatically when $s\rightarrow-s$ due to the symmetry. This in turn means that we will only be interested in $\beta_2$ and  $\beta_3$ in our discussions since $\beta_2>\beta_1$ when $s>0$. It is  important to emphasize that, until this point, we have not paid attention to the internal structure of the green (or blue) triangular region. In fact, in terms of entanglement dynamics, there is no difference between darker or brighter parts in the green (or blue) triangular region. However, the internal regions with different brightness will turn out to be significant for distinguishing different dynamical behaviors of local quantum uncertainty. 

In the upper part of the red triangular region where $r<0.25$ and $s>0$, one can show that $\beta_2$ is always greater than $\beta_3$ during the dynamics and moreover it is monotonically decreasing, which means that the local quantum uncertainty in this region is monotonically increasing. In other words, in the red region, while entanglement remains forever frozen, local quantum uncertainty is smoothly amplified at the same time. An explicit example of this situation is displayed in Fig. \ref{fig2}(a) where a pair of initial states are examined with parameters $r=0.15,s=0.07$ and $r=0.15,s=0.12$. We now turn our attention to different shades of the green triangular region. In the first brightest green part where $r\in(1/3,1/2)$, even though we observe entanglement sudden death, local quantum uncertainty still grows monotonically as $\beta_2>\beta_3$ still holds at all times. We show an instance of this case in Fig. \ref{fig2}(b) where we fix the initial parameters as $r=0.32,s=0.25$ and $r=0.32,s=0.3$ for a pair of states. Proceeding to the next part in the middle of the green triangular region, it is possible to show that the local quantum uncertainty of the initial states from this region initially start to monotonically increase but then at a critical point a sudden change in the dynamics of the local quantum uncertainty occurs and it starts to decay. The reason behind this behaviour is the fact that, despite $\beta_2>\beta_3$ initially in this region, as the system evolves in time, there takes place a transition and $\beta_3$ becomes greater than $\beta_2$. The conditions for the sudden transition to occur (which also defines the borders of the middle region inside the green triangle) is given by the inequalities $3r-s<1$ and $r>1/3$. The sudden transition time can also be calculated as
\begin{equation}
t_{st}=-\frac{1}{2\Gamma} \ln\left[\frac{3r-1}{s}\right].
\end{equation}
We present a clear example of this scenario in Fig. \ref{fig2}(c) for a pair of initial states where we the parameters are set as $r=0.37,s=0.25$ and $r=0.37,s=0.35$. Lastly, we investigate the remaining darkest green part in Fig. \ref{fig1}. In this region, it is easy to show that $\beta_3$ is always greater than $\beta_2$ and furthermore it is monotonically increasing throughout the evolution of the system. Consequently, local quantum uncertainty for the initial states in this region is monotonically lost to a certain degree during the dynamics. Fig. \ref{fig2}(d) displays an example of this situation for a pair of initial states where we fix the parameters as $r=0.45,s=0.2$ and $r=0.45,s=0.35$.

\section{Summary and Conclusion}
\vspace{-0.1cm}
In summary, we have studied the dynamics of quantum entanglement and more general quantum correlations, quantified by negativity and local quantum uncertainty, respectively, under collective pure dephasing noise. We have considered two-qubit quantum systems described by a family of density matrices that are defined with two real valued parameters. Our treatment of the problem has allowed us to express our findings using an illustrative geometric interpretation. Our investigation not only extends the results obtained in Refs. \cite{theo,exp} regarding the phenomenon of time-invariant entanglement but also provides an analysis of the relation of different dynamical behaviors of entanglement to more general quantum correlations quantified via local quantum uncertainty.

First, we have determined the conditions for quantum entanglement to become forever frozen, i.e. time-invariant, during the time evolution of the open system. Moreover, we have also identified the set of initial state parameters for which the initial amount of entanglement in the system is completely due to the sudden death phenomenon. Second, in relation with different regimes of entanglement dynamics, we have explored the evolution of local quantum uncertainty. We have demonstrated that in the regime of parameters where entanglement is time-invariant, it is guaranteed that the local quantum uncertainty is smoothly generated. On the other hand, looking at the region where sudden death of entanglement can be observed, we have determined several different dynamical behaviors for the local quantum uncertainty. For instance, a particularly interesting case is the occurrence of a sudden change in dynamics of the measure which prevents its amplification. All in all, our results show that, in the parameter space of the considered density matrices, entangled initial states with $r<1/4$ can both maintain their entanglement constant and increase their local quantum uncertainty. Besides, entangled initial states with $1/4<r<1/3$ can at least increase their local quantum uncertainty despite the inevitable sudden death of entanglement. Finally, we note that although the considered model here is quite simple and the decay factors can take more complicated forms depending on the physical systems, our outcomes on invariant entanglement would remain the same even for such models, as they are independent of the details of the decay factor. Also, even though the amplification rate or decay rate of local quantum uncertainty are dependent on the specifics of the decay factor, the general trend of the dynamical behavior for the correlations would be similar.

Since quantum entanglement and more general discord-like quantum correlations have the potential to be used as a resource for the possible applications quantum information science, we hope that our results on their protection or generation might be of relevance in related problems.


\begin{thebibliography}{100}

\bibitem{entrev} R. Horodecki, P. Horodecki, M. Horodecki, and K. Horodecki, Rev. Mod. Phys. \textbf{81}, 865 (2009).
\bibitem{book} M.A. Nielsen and I.L. Chuang, \emph{Quantum Computation and Quantum Information}, (Cambridge, Cambridge University Press, 2000).
\bibitem{sep} B. P. Lanyon, M. Barbieri, M. P. Almeida, and A. G. White, \emph{Experimental Quantum Computing without Entanglement}, Phys. Rev. Lett. {\bf 101}, 200501 (2008).
\bibitem{discrev}  K. Modi, A. Brodutch, H. Cable, T. Paterek, and V. Vedral, Rev. Mod. Phys. \textbf{84}, 1655 (2012).
\bibitem{WYSI} E. P.Wigner and M. M. Yanase, Proc. Natl. Acad. Sci. USA \textbf{49}, 910 (1963).
\bibitem{WYSIqc} D. Girolami, T. Tufarelli, and G. Adesso, Phys. Rev. Lett. \textbf{110}, 240402 (2013).
\bibitem{bp} H.-P. Breuer and F. Petruccione, \emph{The Theory of Open Quantum Systems}, (Oxford Univ. Press, 2007). 
\bibitem{sd} T. Yu, J.H. Eberly, Phys. Rev. Lett. \textbf{97}, 140403, (2006) .
\bibitem{robust} T. Werlang, S. Souza, F. F. Fanchini, and C. J. Villas Boas, Phys. Rev. A {\bf 80}, 024103 (2009).
\bibitem{sabri} L. Mazzola, J. Piilo, S. Maniscalco, Phys. Rev. Lett. \textbf{104}, 200401 (2010).
\bibitem{frezunc} B. Aaronson, R. Lo Franco, G. Adesso, Phys. Rev. A \textbf{88}, 012120 (2013).
\bibitem{discrev2} A. Bera, T. Das, D.Sadhukhan, S. S. Roy, A. Sen De, U. Sen, arXiv:1703.10542.
\bibitem{theo} E. G. Carnio, A. Buchleitner, and M. Gessner, Phys. Rev. Lett. \textbf{115}, 010404 (2015).
\bibitem{exp} B.-H.Liu, X.-M. Hu, J.-S. Chen, C. Zhang, Y.-F. Huang, C.-F. Li, G.-C. Guo, G. Karpat, F. F. Fanchini, J. Piilo, S. Maniscalco, Phys. Rev. A \textbf{94}, 062107 (2016).
\bibitem{yueb} T. Yu, J.H. Eberly, Phys. Rev. B \textbf{68}, 165322 (2003).
\bibitem{neg} G. Vidal, R.F. Werner, Phys. Rev. A \textbf{65}, 032314 (2002).
\bibitem{cohh} D. Girolami, Phys. Rev. Lett. \textbf{113}, 170401 (2014).
\bibitem{cohmon} T. Baumgratz, M. Cramer, and M. B. Plenio, Phys. Rev. Lett. \textbf{113}, 140401 (2014).
\bibitem{coh} I. Marvian, and R. W. Spekkens, Nat. Commun. \textbf{5}, 3821 (2014).

\end{thebibliography}
\end{document}